\documentclass[pra,twocolumn,showpacs]{revtex4}
\usepackage{graphicx}
\usepackage{subfigure}
\usepackage[latin1]{inputenc}

\newcommand{\eeq}{\end{equation}}
\newcommand{\beq}{\begin{equation}}

\newcommand{\kq}{$^{41}$K~}
\newcommand{\kt}{$^{39}$K~}

\begin{document}

\title{Intense slow beams of bosonic potassium isotopes}

\author{J. Catani$^{*,1}$, P. Maioli, L. De Sarlo,
F. Minardi$^{1,2}$ and M. Inguscio$^{1,2}$}

\affiliation{LENS - European Laboratory for Non-Linear
Spectroscopy and Dipartimento di Fisica, Università di Firenze,
via N. Carrara 1, I-50019 Sesto Fiorentino - Firenze, Italy\\
$^1$INFN, via G. Sansone
1, I-50019 Sesto Fiorentino - Firenze, Italy\\
$^2$CNR-INFM, via G. Sansone 1, I-50019 Sesto Fiorentino -
Firenze, Italy}

\begin{abstract}
We report on an experimental realization of a two-dimensional
magneto-optical trap (2D-MOT) that allows the generation of cold
atomic beams of $^{39}$K and $^{41}$K bosonic potassium isotopes.
The high measured fluxes up to $1.0\times10^{11}$ atoms/s and low
atomic velocities around 33 m/s are well suited for a fast and
reliable 3D-MOT loading, a basilar feature for new generation
experiments on Bose-Einstein condensation of dilute atomic
samples. We also present a simple multilevel theoretical model for
the calculation of the light-induced force acting on an atom
moving in a MOT. The model gives a good agreement between
predicted and measured flux and velocity values for our 2D-MOT.
\end{abstract}


\date{\today}

\maketitle

\section{INTRODUCTION}

\begin{figure}
\begin{center}
\includegraphics[width=\columnwidth]{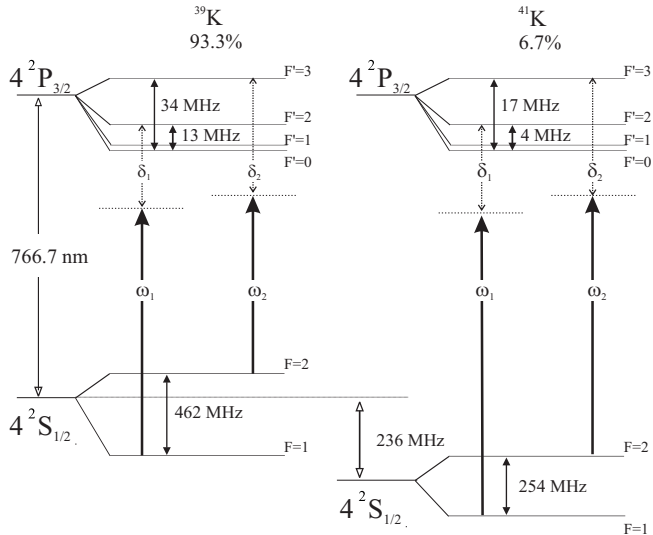}
\caption{Level diagrams for \kt and $^{41}$K. In the upper part of
the diagram is shown the natural abundance of both isotopes.}
\label{klevels}
\end{center}
\end{figure}

In the field of atomic physics, experiments and theoretical
studies on degenerate mixtures of cold gases have been acquiring a
growing interest in the last few years. The vast majority of works
have featured Fermi-Bose mixtures, that have now been realized
with several combinations of atoms: $^6$Li-$^{7}$Li
\cite{salomon}, $^6$Li-$^{23}$Na \cite{ketterledouble},
$^{40}$K-$^{87}$Rb \cite{modugnodouble}, $^6$Li-$^{87}$Rb
\cite{zimmermandouble}. Many experiments focused on the degenerate
Fermi gas, while the bosonic component has only been a tool to
reach Fermi degeneracy via sympathetic cooling. Remarkable
achievements include BCS-BEC crossover \cite{salomonbcs, grimmbcs,
jinbcs, ketterlevortex, dukebcs, huletbcs}, fermionic Bloch
oscillations \cite{modugnointerf}, observation of the 3D Fermi
surface \cite{esslinger}. Besides, Fermi-Bose systems display a
wealth of interesting phenomena genuinely related to the presence
of two species. A few, like interspecies Fano-Feshbach resonances
\cite{jinfeshbach,ketterlefeshbach,lensfeshbach} and boson-induced
collapse of the Fermi gas \cite{collapse-lens, sengstock}, have
already been observed but many more have been proposed and still
await experimental confirmation, as e.g. boson-induced
superfluidity \cite{pethick} and mixed phases in optical lattices
\cite{lewenstein}.\

On the other hand, Bose-Bose mixtures have remained relatively
unexplored: after the pioneering work on $^{41}$K-$^{87}$Rb
\cite{lensk41-rb87}, to our knowledge only one other double-specie
condensate has been recently produced, namely
$^{174}$Yb-$^{176}$Yb \cite{yb-yb-japan}. As for the K-Rb mixture,
the recent precise determination of the interspecies scattering
length by Feshbach spectroscopy \cite{k-rb-feshbachlens} promises
a rich phase-diagram for the two species loaded in an optical
lattice. In view of an experiment on a K-Rb Bose-Bose mixture we
realized and characterized an intense, slow beam of cold bosonic
potassium isotopes.\

An intense and reliable source of cold atoms represents a
favorable starting point for every degenerate mixture experiment.
Two-dimensional magneto-optical traps (2D-MOTs) of rubidium have
been investigated by several groups \cite{walraven,pfau} and
represent one of the brightest sources of slow atoms. With an
appropriate double-chamber vacuum system, the 2D-MOT provides high
atomic flux through a small aperture between the two chambers.
Therefore, it appears perfectly suited for loading an ordinary
3D-MOT in ultra-high vacuum (UHV) environment, the most widely
used pre-cooling stage towards quantum degeneracy of dilute
gases.\

In this work we demonstrate an efficient 2D-MOT for bosonic
potassium isotopes \kt and $^{41}$K, starting from a natural
abundance sample. Previously, only a 2D-MOT of fermionic $^{40}$K
has been implemented \cite{sengstock}, while for bosonic isotopes
no bright sources have been realized. The peculiarity of bosonic
potassium is the tight hyperfine structure of the $D_2$ line
excited state $4 ^2$P$_{3/2}$ \cite{ingusciorevmodphys,sieradzan}:
as shown in Fig.~\ref{klevels}, the separation between hyperfine
levels is comparable to the transition linewidth, $\Gamma=2\pi
\times 6.2$ MHz. This feature represents the main physical
difference with respect to almost all other alkalis (exception
made for Li), and it deeply affects the sub-Doppler cooling
mechanism \cite{bambiniagresti}. The aim of the present work is to
address the question of how efficiently 2D-MOT schemes work for
bosonic potassium.

The paper is organized as follows. In Sec.~\ref{Sec:setup} we
describe the experimental setup and some experimental techniques
used to generate and monitor the cold atomic beam. The results of
the characterization of the 2D-MOT are reported in
Sec.~\ref{Sec:results}, while a theoretical model giving a good
description of data is explained in Sec.~\ref{theoreticalmodel}.
In Sec.~\ref{Sec:conclusions} we conclude with some brief
considerations in view of future experiments.

\section{Experimental Setup}
\label{Sec:setup}
\subsection{Vacuum system}

\begin{figure}
\begin{center}
\includegraphics[width=\columnwidth]{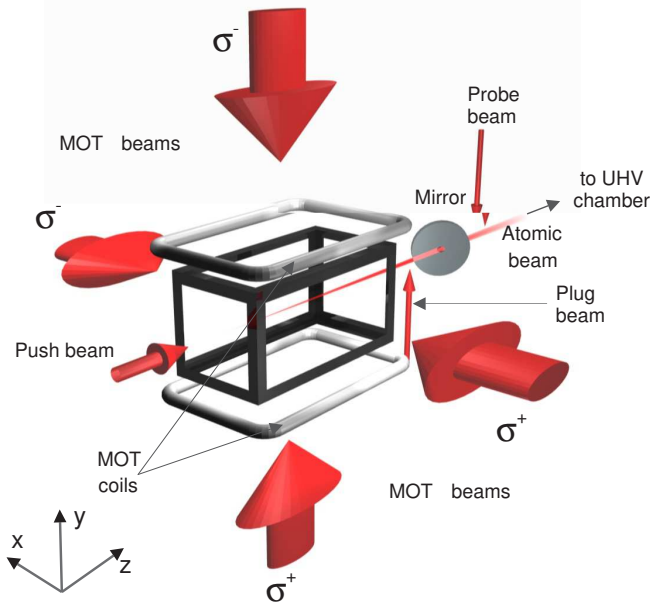}
\caption{(Color online) Schematics of the 2D-MOT system. Two
retroreflected transverse cooling beams, with 3:1 elliptic
section, cross in the center of a Ti structure. A pair of copper
wired coils generates the quadrupole magnetic field. Near the end
of the cell a drilled mirror is placed, tilted 45° with respect to
the longitudinal axis $z$, along which a 1.5 mm waist push beam is
propagating, forcing atoms to travel through the hole. A plug beam
is placed near the hole in the mirror, and a probe beam is used to
detect the atomic flux in the UHV chamber.} \label{scheme}
\end{center}
\end{figure}

The vacuum system consists of two chambers, that we call 2D-MOT
and UHV chamber. The 2D-MOT apparatus is schematically reported in
Fig.~\ref{scheme}. Four rectangular 80$\times$35 mm BK7
anti-reflection coated windows, providing optical access to four
transverse trapping beams, are glued on a metallic rectangular
frame, machined from a titanium block. The atomic vapor is
released in natural abundance by dispensers, whose injection
current controls the pressure in the 2D-MOT chamber. The vacuum
setup design gives the possibility to connect to the UHV chamber
another identical 2D-MOT chamber. In this scheme it is possible to
load a Rb-K mixed 3D-MOT starting from two independently
controllable atomic beams.

A flexible, 10 cm long, stainless steel bellow connects the 2D-MOT
and the UHV chambers. A metallic mirror, tilted 45$^\circ$ with
respect to the longitudinal axis, is placed near the end of the
2D-MOT chamber and close to the bellow. A 1-mm hole is drilled at
the center of the mirror to let the atoms exit the 2D-MOT chamber.
This mirror allows us to send a hollow beam, counterpropagating
with respect to the atoms, so that we can turn to a 2D$^+$-MOT
configuration \cite{walraven}.

The 2D-MOT chamber is pumped by a 20 l/s ion diode pump, while in
the UHV chamber a 55 l/s ion pump, combined with the tiny
conductance of the mirror hole, generates a pressure ratio between
the two chambers up to 10$^{-4}$. Inside the bellow, three
graphite tubes of increasing inner diameter (6 to 10 mm) improve
the capability of pumping alkali metals.

The quadrupole magnetic field is generated by a pair of
90$\times$40 mm rectangular coils, fixed around the protruding
glass windows. This elongated shape yields the required almost 2D
axial symmetry to the trapping field. The magnetic gradient in the
$x-y$ plane is set to 17 Gauss/cm. The current in each coil is
independently adjusted to compensate for stray magnetic fields
along the vertical $y$ direction. The compensation of horizontal
fields along $x$ is provided by a single large coil, placed far
from the chamber. In this configuration, the longitudinal magnetic
gradient along $z$ is a factor of 10 smaller than in the $x-y$
plane and it will be hereafter neglected.

\subsection{Laser system}

The cooling and trapping of alkali atoms requires in most cases
two laser frequencies which are conventionally called
\textit{repumping} and \textit{cooling}. The force exerted upon
the atoms is dominated by the cooling light, tuned to a closed
transition, whereas the repumping light is mainly needed to
counteract the hyperfine optical pumping in the ground state,
consequent to off-resonant excitation. As such, only a small
fraction of the laser power suffices for the repumping light.

The situation is completely different for bosonic potassium, where
the tight hyperfine $4 ^2$P$_{3/2}$ level spacings are comparable
to the natural linewidth $\Gamma $. In this case, the cooling
transition $|F=2 \rangle \rightarrow | F'=3\rangle $ is not closed
since the $|F'=1,2\rangle $ states are excited with similar rates.
A fast depletion of the $|F=2\rangle$ ground state toward the
$|F=1\rangle$ ensues, the repumping light needs to be fairly
intense and the cooling force arises from both frequencies
\cite{fortbambini}. Hence, the distinction between cooling and
repumping light makes no sense. However, according to a widespread
convention, we will also call {\it repumping} and {\it cooling}
the transitions indicated respectively by $\omega_1$ and
$\omega_2$ in Fig.~\ref{klevels}.

The $\omega_1$ light is generated by a commercial high-power laser
diode with back-emission stabilization grating, providing 400 mW
at 767.5 nm. The frequency stabilization is achieved by locking
this laser using modulation-transfer saturation spectroscopy in a
reference glass cell containing potassium vapor. To generate the
$\omega_2$ radiation, a beam of about 40 mW is split from the
laser output, frequency shifted by an acousto-optic modulator
(AOM) and then injected into a semiconductor tapered amplifier
that delivers 900 mW. In this way, only the repumping laser needs
to be frequency locked and the frequency difference between the
two sources is precisely set by the radio-frequency signal fed
into the AOM. Each of the two main beams is split into several
beams with independent intensity and frequency control, achieved
by means of double passage AOM stages.
\begin{figure}
\begin{center}
\includegraphics[width=0.7\columnwidth]{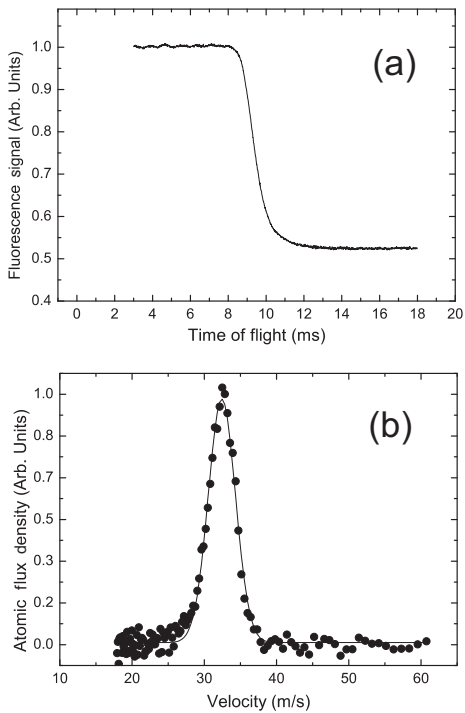}
\caption{(a) Typical acquisition of the fluorescence signal. At
t=0 ms the plug beam is turned on, and after a delay of $\sim 8$
ms the fluorescence starts decaying to zero. \quad(b) Measured
velocity distribution (dots) of atomic beam obtained by a discrete
derivative of (a) and Gaussian fit to data (solid line). The
fitted peak velocity is 32 m/s with a FWHM of 4.5 m/s.}
 \label{tof}
 \end{center}
\end{figure}

After a fiber mode-cleaning stage, a pair of cylindrical lenses
gives a 3:1 elliptical shape, with a smaller waist of 9.4 mm, to
two 2D-MOT transverse beams, which are circularly polarized and
retroreflected after the chamber. The maximum laser power for each
transverse beam is 50 mW and 80 mW for $\omega_1$ and $\omega_2$
respectively.

Along the $z$ axial direction, we shine an additional $\sigma^+$
polarized beam, called \textit{push}, with a waist of 1.5 mm. To
achieve the 2D$^+$-MOT configuration, a fraction of this beam is
split and directed along the $-z$ direction (\textit{slowing}
beam). We recall that the slowing beam has a hollow profile since
it enters the 2D-MOT after reflection upon the drilled mirror.

Another vertical 2 mW beam, called \textit{plug}, resonant with
the cooling transition, intercepts the atomic flux near the mirror
hole: atoms travel unperturbed to the UHV chamber when the plug
beam is off and are swept away when the plug beam is on
\cite{metcalf}. We use this beam to perform time-of-flight (TOF)
measurements, as described below.

\subsection{Detection technique}

As in Ref. \cite{walraven}, we analyze the atomic flux by means of
TOF fluorescence detection. This is accomplished exposing the
atomic flux to a vertical sheet of light (\textit{probe} beam), 30
cm after the mirror hole, and imaging the emitted fluorescence
into a broad area photodiode. The peak intensity of the probe beam
exceeds 0.5 W/cm$^2$, divided between $\omega_1$ and $\omega_2$,
both resonant, in a ratio 1:2.

Switching on the plug beam we interrupt the atomic flux; then,
from the analysis of the decaying fluorescence signal $S_F(t)$ as
a function of time, we obtain the longitudinal velocity
distribution $\rho(v)$ of atoms. Denoting with $\tau$ the time
required for an atom with velocity $v$ along $z$ to travel the
distance $L$ between the plug beam and the probe light sheet, we
can write the total flux $\Phi$ as:
 \beq
 \Phi=\int{\rho(v)dv},
\quad\mathrm{where}\quad\rho(v)=-\frac{\tau}{\eta}\frac{d}{d\tau}S_{F}(\tau)
,\label{velocitydistribution}
 \eeq
with $\tau=L/v$. Here $\eta$ is an experimental coefficient
accounting for the calibration of the photodiode and for the
collection solid angle.
\begin{figure*}
\begin{center}
\includegraphics[width=1.4\columnwidth]{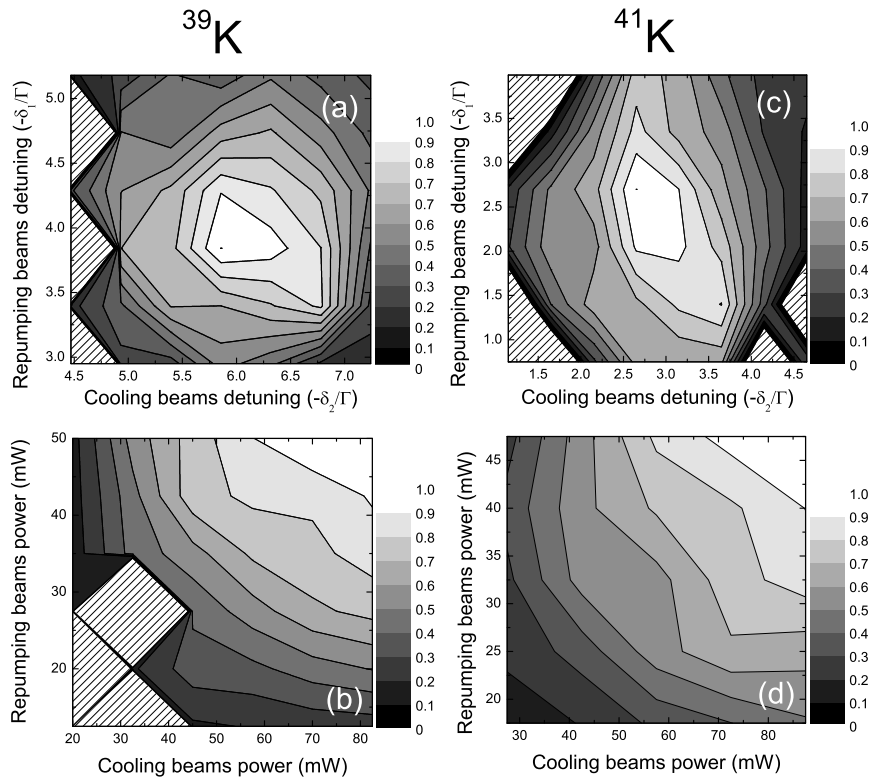}
\caption{Measured \kt atomic flux normalized to unity as a
function of transverse 2D-MOT beams detuning (a), and of their
intensities (b). The detuning values that maximize the atomic flux
are $\delta_1=-3.9\Gamma$ and $\delta_2=-5.8\Gamma$. In (c) and
(d) we report equivalent results obtained for \kq atomic beam, for
which the optimal detunings are $\delta_1=-2.5\Gamma$ and
$\delta_2=-3\Gamma$. No data are taken in the patterned regions.}
\label{k39detuningmatrix}
\end{center}
\end{figure*}

Fig.~\ref{tof}(a) shows a typical acquisition of the TOF signal,
while the corresponding velocity distribution $\rho(v)$, as
derived from Eq.~(\ref{velocitydistribution}), is displayed in
Fig.~\ref{tof}(b). To obtain the total atomic flux $\Phi$, the
peak velocity and the velocity spread, we fit the data with a
Gaussian function.

\section{Experimental results}
\label{Sec:results}

We report the measured flux and velocity distribution of \kt and
\kq atomic beams as a function of experimental parameters. We
shifted from the first to the second isotope by changing the AOMs
frequencies and the laser frequency main locking point, preserving
the geometrical alignment of the 2D-MOT beams. The analysis on \kq
was started only after optimization of frequencies and power for
the $^{39}$K, the former having lower relative abundance ($6.7\%$)
than the latter ($93.3\%$).

The region of vapor pressure spanned in the experiment goes from
$2.9\times10^{-8}$ to $3.9\times10^{-7}$ mbar, corresponding to
the maximum allowed current in the dispenser. We estimate the
background gas pressure to be around 10$^{-9}$ mbar. If not
otherwise specified, in the experiment we set the K gas total
pressure to 7$\times 10^{-8}$ mbar.

The divergence of the atomic beam is measured by imaging with a
CCD camera the fluorescence emitted in the $-x$ direction. The
image profile along $z$ corresponds to the Gaussian profile of the
short axis of probe light sheet, while the vertical ($y$
direction) image profile extension is limited by the mirror hole.
From the known width of the probe beam we calibrate the image
magnification and therefore measure the size of the atomic beam in
the $y$ direction. Given the distance of the probe beam from the
mirror hole, we calculate a divergence of ($34\pm6$) mrad.

\subsection{Trapping beam parameters}

To characterize the 2D-MOT, we first set the intensity of both
repumping and cooling 2D-MOT beams to their maximum, respectively
50 and 80 mW per beam. We then make a scan of both frequencies
$\omega_1$ and $\omega_2$, searching for the values that maximize
the fluorescence signal, hence the total flux of atoms.

We only present data on the total flux $\Phi$ because the atomic
velocities display no significant variations with the 2D-MOT
parameters. The measured peak velocities span a range between 28
and 35 m/s, while the typical distribution spread is 4.5 m/s
(FWHM). Experimental results for \kt detunings are plotted in
arbitrary units normalized to the maximum value in
Fig.~\ref{k39detuningmatrix}(a). The detunings are defined
throughout as follows: $\delta_1=\omega_1 - \omega_{12}$ and
$\delta_2=\omega_2-\omega_{23}$, where we denote with
$\omega_{FF'}$ the atomic transition
$|4^2$S$_{1/2},F\rangle\rightarrow |4^2$P$_{3/2},F'\rangle$. Note
that for these measurements the push beam contains only the
repumping component with a power of 6 mW and frequency
$\omega_p=\omega_{12}-5.2\Gamma$. A more detailed analysis of the
push beam features is reported in Sec.~\ref{push}. The detunings
optimizing the atomic flux are $\delta_1 = -3.9\Gamma$ and
$\delta_2 = -5.8 \Gamma$, which correspond to neither of the two
configurations reported in \cite{fortbambini}. In that work, a MOT
is best loaded with both lasers detuned below all hyperfine
components, in a Raman configuration such as $|\delta_2 - \delta_1
| = \Delta_{32}$ where $\Delta_{32}=2\pi\times 21$ MHz is the
hyperfine separation between the $|F'=3\rangle$ and $|F'=2\rangle$
excited levels. Similar results were later confirmed in
\cite{prevedelli}. We speculate that, since our available laser
beam intensities are a fifth to a tenth of those in
\cite{fortbambini}, we are unable to reach such large values of
detunings.

Then we fix the detunings and decrease both the cooling and
repumping beams intensity. Fig.~\ref{k39detuningmatrix}(b) shows
the \kt atomic flux as a function of the beams intensity: as we
see, there are no maxima in the explored range. Thus, an increase
of the laser power should make the atomic beam more intense.

The analysis on $^{41}$K is performed along the same lines. Here,
the push beam power is set to 6 mW and its frequency to
$\omega_p=\omega_{12}-4.5\Gamma$. Experimental results are plotted
in Fig.~\ref{k39detuningmatrix}(c, d). As before, while a peak is
clearly visible in the frequency dependence of the signal, there
is none in the intensity-dependent plot. Even for $^{41}$K, more
laser power would enhance the number of atoms in the beam.

The largest contribution to the error on the measured total flux
comes from the calibration parameter $\eta$, more precisely from
the evaluation of the solid angle in which fluorescence is
collected. We estimate this systematic uncertainty to be 20\%,
while the statistical error is 5\%.

\begin{figure}
\begin{center}
\includegraphics[width=0.8\columnwidth]{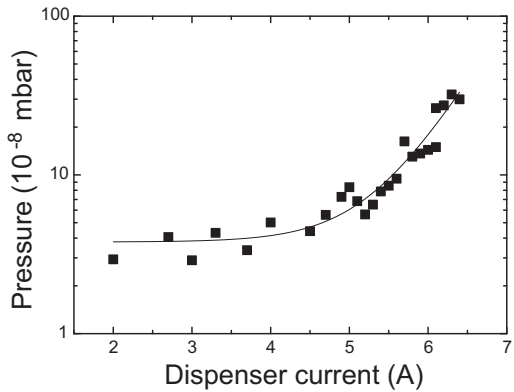}
\caption{Measured \kt gas pressure $p$ in the 2D-MOT chamber
reported as a function of dispenser current $I$. The solid line is
a fit obtained using Eq.~(\ref{press-i-eq}).} \label{fitpress}
\end{center}
\end{figure}

\begin{figure}
\begin{center}
\includegraphics[width=0.8\columnwidth]{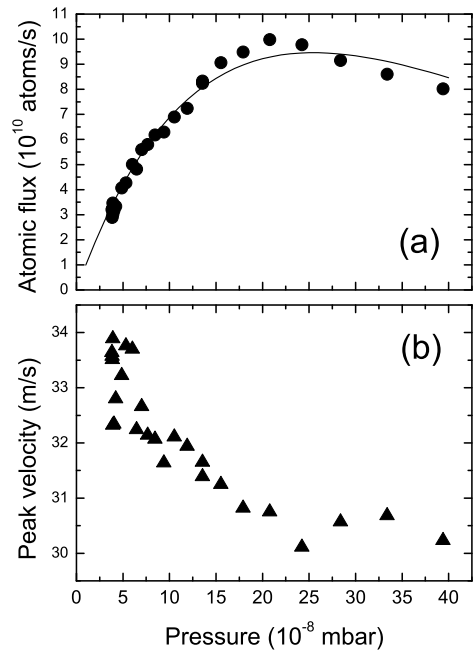}
\caption{Measured \kt atomic flux (a) and peak velocity (b)
reported as a function of gas pressure in the 2D-MOT chamber. Over
the critical value $p=2.1\times 10^{-7}$ mbar the atomic flux
starts to be depleted by collisional effects. The solid line is a
fit obtained using Eq.~(\ref{tcool-coll}).} \label{fluxvelvspress}
\end{center}
\end{figure}

\subsection{Vapor pressure}
\label{pressure}

In order to investigate the dependence of the atomic flux upon the
potassium gas pressure, we found it convenient to establish a
calibration of the vapor pressure $p$ against the dispenser supply
current $I$. This is done by probing the atomic vapor using linear
absorption spectroscopy. We simultaneously record the Doppler
absorption profiles of the cooling transition in the 2D-MOT
chamber and in the reference cell used for laser frequency
locking. The pressure of the latter is inferred from its
temperature: $2.2\times10^{-7}$ mbar at 43 °C. The calibration
curve of $p$ as a function of $I$ is well-fit by an exponential
function:
 \beq
 p= A + \exp(I/I_0-B),
 \label{press-i-eq}
 \eeq
with $A=(3.8\pm1)\cdot10^{-8}$ mbar, $I_0=(0.54\pm0.07)$ A and
$B=(26.7\pm1.8)$. Both the experimental data and the fit for \kt
are reported in Fig.~\ref{fitpress}.

The maximum allowed current in the dispenser, 6.4 A, sets the
highest pressure we reach, i.e. $3.9\times10^{-7}$ mbar. Below the
dispenser ignition point, a residual pressure of $4\times10^{-8}$
mbar is likely due to potassium vapor slowly released by the
chamber walls.

In Fig.~\ref{fluxvelvspress}(a) we report the measured atomic flux
as a function of the pressure for $^{39}$K. As we can see, the
flux increases with pressure until a critical point and then
decreases. In the case of \kt this critical point occurs at
$p=2.1\times 10^{-7}$ mbar where a flux of $1.0\times 10^{11}$
atoms/s is reached. Such a critical point has been observed also
for Rb \cite{walraven,pfau, dalibard} and ascribed to collisions
with the background vapor. When the inverse collision rate becomes
of the same order of the cooling time $t_c$, i.e. the time
required to reach the mirror hole, the probability for an atom to
reach the mirror hole drops. Assuming this is the case also for
$^{39}$K, we fit our data to obtain the collision rate. The total
flux is: \beq \Phi = \int_0^{\infty}{\varphi(t_c) dt_c}
\label{eqflussoint} \eeq where $\varphi(t_c)$ is the cooling time
distribution.

In the collisionless regime $\varphi(t_c)$ is linearly
proportional to the density, hence the pressure, in the 2D-MOT
chamber. In presence of collisions with the background vapor,
$\varphi(t_c)$ is depleted for $t_c > \gamma$, where
$\gamma=\kappa p$ is the collision rate, proportional to the
pressure $p$. Indeed, a single collision with an atom at room
temperature ($v_{th}=250$ m/s) is sufficient to remove the cold
atoms from the 2D-MOT velocity capture range. Then, if we assume
that the cooling time distribution is a Gaussian
\[
\varphi(t_c)=\frac{\Phi}{\sqrt{2\pi}\sigma}
\exp\left(-\frac{(t_c-t_0)^2}{2 \sigma^2}\right),
\]
Eq.~(\ref{eqflussoint}) is modified to
\begin{eqnarray}
\label{tcool-coll}
\Phi &=& \int_0^{\infty}{\varphi(t_c) e^{-\gamma t_c} dt_c} \nonumber \\
&=& \frac{\Phi}{2}e^{- \gamma t_0 + \gamma^2 \sigma^2/2}\left(1 +
{\rm Erf}\left(\frac{t_0 - \gamma \sigma^2}
{\sqrt{2}\sigma}\right)\right),
\end{eqnarray}
where ${\rm Erf}(x)=\frac{2}{\sqrt\pi}\int_0^x{\exp(-t^2)dt}$ is
the usual definition of the error function.\

From the numerical simulations illustrated below, we obtain the
first and second moment of the cooling time distribution:
$t_0=5.1$ ms and $\sigma=1.7$ ms. We then use
Eq.~(\ref{tcool-coll}) to fit the data, with two free parameters:
an overall scaling factor and the collision rate to pressure ratio
$\kappa$. The results of the fitting is shown by the solid line in
Fig.~\ref{fluxvelvspress}(a). The collision rate given by the fit
$\gamma$(s$^{-1})=(8.6\pm 0.6)\times p~(10^{-8}$ mbar) is
approximately one order of magnitude larger than the value
reported in \cite{walker}: 0.3 s$^{-1}$ at $p=4\times
10^{-9}$~mbar, i.e. $0.75$ s$^{-1}\times p~(10^{-8}$ mbar). Our
value is closer to the collision rates observed with rubidium. A
meaningful comparison with Ref. \cite{walker} requires a more
detailed investigation on the role of light-assisted collisions,
assumed negligible for the 2D-MOT.\

Another noticeable effect upon pressure increase is the slowing
down of the atomic beam, as shown by the plot of the peak velocity
in Fig.~\ref{fluxvelvspress}(b). This behavior is in contrast with
the case of the simple rubidium 2D-MOT described in \cite{pfau},
where an increase of the gas pressure corresponds to the increase
of the peak velocity. Therefore we attribute the observed effect
to the progressive absorption of the push beam along its
propagation path by the atomic gas since, as reported below, the
peak velocity decreases as the push beam intensity is reduced.

As mentioned before, when not otherwise specified all data
reported in this work have been taken at $p=7\times10^{-8}$ mbar.

\subsection{Push beam}
\label{push}

We turn to the experimental investigation of the atomic beam
behavior when the push beam parameters are changed. This is a
crucial characterization since we observe no atomic flux in
absence of the push beam, in accordance with the findings in
experiments with rubidium \cite{walraven, pfau} where the 2D-MOT
without push beam is not efficient in our pressure range.
Surprisingly, for potassium the push beam works best if one uses
only repumping light, and therefore we define the push beam
detuning $\delta_p=\omega_p-\omega_{12}$. We defer the discussion
of this point to the end of the Section.

\begin{figure}[t]
\begin{center}
{\includegraphics[width=0.7\columnwidth]{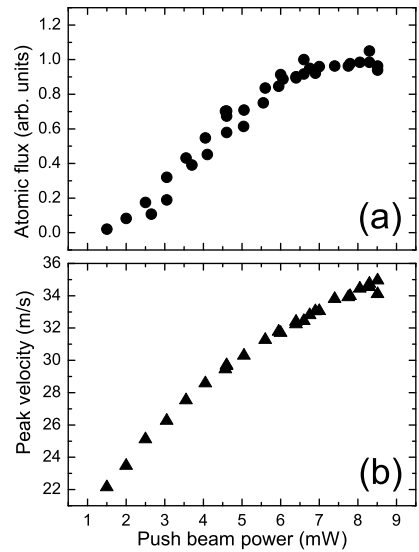}}
\caption{Atomic flux $\Phi$ (a) and peak velocity (b) measured for
\kt atomic beam as a function of the push beam power. In this
analysis we set $\delta_p=-5.2\Gamma$.} \label{k39pushpower}
\end{center}
\end{figure}

First we report the atom flux and peak velocity,
Figs.~\ref{k39pushpower}(a) and \ref{k39pushpower}(b)
respectively, for \kt at fixed detuning $\delta_p=-5.2\Gamma$ as a
function of power. Above 6 mW, the atom number is approximately
constant, while the peak velocity keeps increasing. Therefore, we
use a power equal to 6 mW (peak intensity of 170 mW/cm$^2$) in
order to have the maximum flux with a velocity distribution still
within the capture range of the 3D-MOT in which the atoms will be
collected.

\begin{table*}
\caption{Optimal experimental beams power values (\textit{P}) and
detunings ($\delta$) used to generate a cold beam of \kt and \kq
atoms. In the right part of the table the total flux $\Phi$ values
and the corresponding peak velocities are reported. The pressure
working point is set to $7\times10^{-8}$~mbar. The magnetic field
gradient is 17 Gauss/cm.}
\begin{ruledtabular}
\begin{tabular}{lcccccc|cc}
&\multicolumn{2}{l}{~~Repumping Beams}&\multicolumn{2}{l}{~~~~Cooling Beams}&\multicolumn{2}{l|}{~~~~~Push Beam}&$\Phi$&Peak Velocity\\

&\textit{P}~[mW]&$\delta_1$~[$\Gamma$]&\textit{P}~[mW]&$\delta_2$~[$\Gamma$]&\textit{P}~[mW]&$\delta_p$~[$\Gamma$]&[atoms/s]&[m/s] \\

\hline
&&&&&&&&\\
{\bf \kt} &50&-3.9&80&-5.8&6&-5.2&$6.2\times10^{10}$&35
\\
&&&&&&&&\\
{\bf \kq} &47&-2.5&85&-3.0&6&-4.5&$5.2\times10^{9}$&33\\

\end{tabular}
\end{ruledtabular}
\label{tabella}
\end{table*}

No clear enhancements of the flux stem from the addition of a
counterpropagating hollow beam along $-z$ direction (2D$^+$-MOT
configuration). Only a 20\% reduction in mean velocity was found
for a $\sigma^-$ beam polarization.

Given the hyperfine structure of potassium, one would expect to
increase the efficiency of the push force using both cooling and
repumping beams, since this avoids the depletion of the ground
state population. On the contrary, the atomic beam is deteriorated
even by a small fraction of cooling light $\omega_2$ and it is
almost extinguished when the $\omega_2$ intensity approaches that
of $\omega_1$. We attribute this effect to an increase of both the
longitudinal velocity and the radial temperature of the atoms,
when pushed by both frequencies. Instead, with a single frequency
the hyperfine optical pumping confines the action of the push
force within the 2D-MOT volume and the atoms drift freely after
the mirror hole.

\begin{figure}[b]
\begin{center}
\includegraphics[width=\columnwidth]{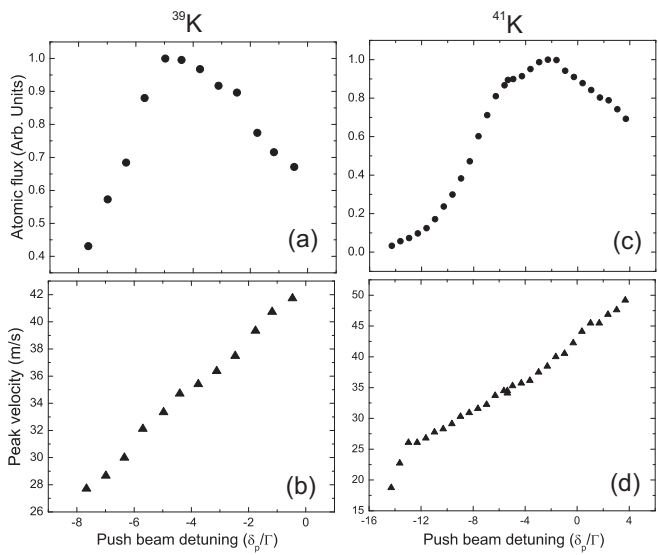}
\caption{Measured atomic flux $\Phi$ and peak velocity of \kt (a,
b) and \kq (c, d) as a function of the push beam frequency
detuning $\delta_p$. The push beam power is set to 6~mW.}
\label{pushfreq}
\end{center}
\end{figure}
One may argue that the push beam should then work also with the
frequency $\omega_2$ alone. We experimentally find that, at equal
power, $\omega_2$ light is far less efficient than $\omega_1$. We
speculate that this is due to $\omega_2$ being more prone to
hyperfine optical pumping for two reasons: \textit{(i)} in the
2D-MOT there is more cooling than repumping light, \textit{(ii)}
$\omega_2$ detuned a few linewidths to the red of the $|F=2\rangle
\rightarrow |F'=3\rangle$ transition is near to the $|F=2\rangle
\rightarrow |F'=1,2\rangle$ resonances causing hyperfine optical
pumping (the same effect is less important for the $\omega_1$
light, since the transitions $|F=1\rangle \rightarrow
|F'=2,1\rangle$ are closer and there is no hyperfine optical
pumping for $|F=1\rangle \rightarrow |F'=0\rangle$). Another
possible way to avoid hyperfine optical pumping is to use a blue
detuned cooling light \cite{swanson}, even if this would increase
the pushing efficiency mainly on fast atomic classes and hence
confer to the atomic flux a higher peak velocity.

As Fig.~\ref{k39pushpower}(a) shows, decreasing the intensity of
the laser beam the atomic flux decreases as well. In the pure
2D-MOT configuration, when no push beam is shone on atoms, no flux
is detected, independently of both power and frequency of
transverse beams and quadrupole magnetic field gradients.

\subsection{Summary}

In Tab.~\ref{tabella} are summarized all the experimental
parameters which maximize the total atomic flux $\Phi$ of the two
isotopes and the values of the corresponding fluxes. For
$p=7\times10^{-8}$ mbar high atomic fluxes are achieved,
containing $6.2\times10^{10}$ atoms/s and $5.2\times10^9$ atoms/s
for \kt and \kq respectively. The pressure value and their small
peak velocities, respectively 35 and 33 m/s, are perfectly
adequate to an efficient subsequent loading in a UHV environment
3D-MOT. By increasing $p$ up to $2.1\times10^{-7}$ mbar a flux of
$1.0\times 10^{11}$ atoms/s of \kt is then observed. Nearly the
same gain is expected for \kq flux as well at the same value of
pressure $p$.

The high intensity of atomic fluxes is an essential point in order
to reach a degenerate regime for K isotopes following a double MOT
cooling and trapping scheme. At optimal detunings, with a 15 times
lower light intensity, we achieve a 3D-MOT loading rate higher
than the one reported in \cite{fortbambini}. Thus, we conclude
that, to generate a cold atomic beam, a 2D-MOT shows a much higher
efficiency when compared to an ordinary 3D-MOT.

\section{A simplified multilevel theoretical model}
\label{theoreticalmodel}

To have a quantitative description of our observations we use an
extension of the theoretical model discussed in \cite{dalibard}.
This model shortcuts the integration of the optical Bloch equation
by assuming a heuristic expression of the total force exerted by
the different beams, all having the same frequency, on a two-level
atom:

 \begin{eqnarray}
 \label{forcedalibard}
 {\bf f}&=&\frac{\hbar\Gamma}{2}\sum _{i}{\bf k}_i
 \frac{s_i}
 {1+\sum_{\scriptstyle j}s_j},\\
 s_j&=&\frac{I_j}{I_s}\frac{\Gamma^2}{\Gamma^2 + 4(\delta_j - {\bf k}_j\cdot {\bf v})^2} \nonumber
 \end{eqnarray}
where $i,j$ denote the beams and $I_s=\pi hc \Gamma/(3 \lambda^3)$
is the two-level saturation intensity, equal to 1.8 mW/cm$^2$ for
potassium $D_2$ line. Authors of Ref.~\cite{dalibard} employ this
model to analyze a 3D-MOT of Rb with the addition of a push beam.
The extension of this treatment to bosonic potassium, because of
the narrow upper level structure, requires to take into account
all the allowed hyperfine transitions. For this purpose, we
introduce the further assumption that forces arising from
different transitions add independently. This consistently
disregards coherences among the two 4S$_{1/2}$ hyperfine states,
which however play no role in the Doppler cooling mechanism. In
principle, we should consider even the Zeeman structure of the
hyperfine levels; in practice, to reduce the number of transitions
contributing to the total force, we calculate the detunings and
the line strength in a manner to average out the Zeeman sublevels.
For each laser beam $i$ and each transition $|F\rangle \rightarrow
|F'\rangle$, we define the average detuning $\Delta^{FF'}_i$ as
the center-of-mass of all the Zeeman components weighted by the
squared Clebsch-Gordan coefficients $|c(F',m';1,\sigma_i,F,m)|^2$,
where $\sigma_i$ denotes the beam polarization (all beams are
circularly polarized with the MOT required helicity). This
detuning depends linearly on the displacement from the 2D-MOT axis
via the magnetic field gradient. We then define the strength of
each hyperfine transition:

\[
G^{FF'}_i=\frac{\sum_{m,m'}
|c(F',m';1,\sigma_i,F,m)|^2}{\sum_{F',m,m'}
|c(F',m';1,\sigma_i,F,m)|^2}.
\]

We incorporate hyperfine optical pumping by breaking the force
into two parts, due to the cooling and the repumping light,
weighted by the relative populations in the $F=2$ and $F=1$ ground
levels. Therefore, the expression (\ref{forcedalibard}) of the
total force is generalized as follows:

 \begin{eqnarray}
 \label{totalforce}
 {\bf f}&=&\frac{p_1}{p_1+p_2}{\bf f}_1+\frac{p_2}{p_1+p_2}{\bf f}_2\\
 {\bf f}_F&=&\frac{\hbar\Gamma}{2}
 \sum_i{\bf k}_i
 \frac{s_{{\scriptscriptstyle i},{\scriptscriptstyle F}}}
{1+s_F},\quad F=1,2
\end{eqnarray}

with
 \begin{eqnarray*}
 \label{satparameter}
 s_{i,F}&=&\frac{I_{i,F}}{2I_s} \sum_{F'} G_i^{FF'}
 \frac{\Gamma^2}{\Gamma^2 +4\left(\delta_{i,F}-
 {\bf k}_i \cdot{\bf v}-\Delta_i^{FF'}\right)^2},\\
\end{eqnarray*}
where $s_F=\sum_{j} s_{j}$. Here, $I_{i,{\scriptscriptstyle
1}},I_{i,{\scriptscriptstyle 2}}$ are respectively the repumping
and cooling intensity of the $i$-th beam, and $\delta_1,\delta_2$
the corresponding detunings as defined earlier. The populations
$p_1, p_2$ are taken as the equilibrium values of the rate
equations for the six hyperfine components.

\subsection{Numerical Simulations}
A numerical integration of the classical equations of motion
yields the phase-space trajectories. We consider only the atoms
that, at $t=0$, lie on the boundary surface $S$ of the 2D-MOT
volume, approximated by a rectangular box with sizes equal to the
beam diameters. In sampling the velocity-space, the Boltzmann
factor is nearly unity for all velocities lying within the 2D-MOT
capture range, well below the 250 m/s thermal velocity. From the
integration we extract: \textit{(i)} the fraction $\wp$ of the
atoms exiting the mirror hole to the atoms entering the 2D-MOT
volume, \textit{(ii)} the longitudinal velocity distribution of
the atomic beam, and \textit{(iii)} the distribution of the
cooling time $\varphi(t_c)$, as defined earlier.

To obtain the total flux, we only need to multiply $\wp$ by the
total number of atoms entering the cooling volume per second, at
pressure $p$ and room temperature $T$:
 \beq
 \label{calculatedflux}
\Phi=\wp\times S\frac{p}{\sqrt{2\pi mk_BT}}. \eeq

In our simulation multiple scattering of light and intra-beam
atomic collisions are neglected. The collisions with background
gas, occurring at rate $\gamma=60$~s$^{-1}$, are accounted for by
weighting each trajectory with a factor $\exp(-\gamma t_c)$ to
deplete the tail of atoms with long cooling times $t_c$. We also
select only those atoms flying in a cone 34 mrad wide around the
longitudinal axis $z$.

Setting the experimental parameters as in Table~\ref{tabella},
with a quadrupole magnetic field gradient of 17 Gauss/cm, for \kt
we obtain a total flux $\Phi_{\rm sim}= 8.7\times
10^{10}$~atoms/s, with a peak velocity of 30.2 m/s, in good
agreement with the experimental values reported in
Table~\ref{tabella}.

\begin{figure}
\begin{center}
\includegraphics[width=\columnwidth]{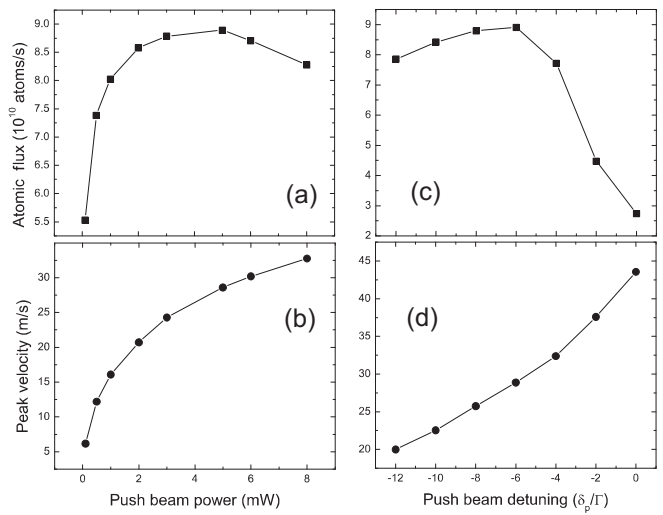}
\caption{Results of simulations for \kt: total flux and peak
velocity as a function of the push beam power (a,b) and detuning
(c,d).} \label{sim-plots}
\end{center}
\end{figure}
Repeating the simulation for different values of the push beam
power and detuning we find the curves plotted in
Fig.~\ref{sim-plots}. The agreement with the experimental findings
is satisfactory for the peak velocities. As for the total flux,
the model shows saturation in push power at lower values than in
the experiment and an optimum detuning of $-6\Gamma$ close to the
observed value of $-5.2 \Gamma$. The calculated dependence of
atomic flux on push beam power and detuning reproduces only
qualitatively the experimental curves of Figs.~6(a) and 7(a). We
believe that coherent effects ignored in our simulations are
likely responsible for the discrepancies with data. A more exact
analysis based on the integration of optical Bloch equations is
needed to address these issues but is beyond the scope of our
simplified model.

\section{Conclusions and outlook}
\label{Sec:conclusions} The main conclusion of this work is that,
similar to rubidium, potassium atoms can be coaxed into an intense
and cold beam, still preserving the UHV constraints. Recently, we
have used two separate 2D-MOT beams to load a double-specie K/Rb
MOT in the UHV chamber as a preliminary step towards quantum
degenerate Bose-Bose mixtures. The combination of potassium and
rubidium offers a wealth of possibilities to tune the interspecies
interactions via Feshbach resonances \cite{k-rb-feshbachlens}.
Broad Fano-Feshbach resonances are predicted for
$^{39}$K-$^{87}$Rb and $^{41}$K-$^{87}$Rb in the absolute ground
state at 318 G and 515 G, respectively. For $^{39}$K, sympathetic
cooling with Rb can be a route to single condensates with tunable
interactions, where resonances are expected around 40 G for two
atoms in the magnetically trapped state $|F=1,m=-1\rangle$
\cite{bohn-feshbach39} and around 20 G for the absolute ground
state \cite{tiesinga}.

An exciting perspective is to tune interactions in the presence of
an optical lattice, where a rich diagram of insulating and
superfluid phases is predicted when the interspecies scattering
length is varied \cite{two-species-lattice-liu}. Also, as
suggested in \cite{mott-feshbach-molecules}, entering the
Mott-insulator regime for both species paves the way to
Feshbach-associated long-lived bosonic heteronuclear molecules,
since three-body losses can be suppressed by tailoring the Mott
phase to
have one atom per species in each lattice site.\\

This work had financial support from Ente Cassa di Risparmio in
Firenze and from Istituto Nazionale di Fisica Nucleare through
project SQUAT. We warmly thank P. Clad\'e for his help, G. Modugno
for many fruitful suggestions, C. Fort and all the members of the
Quantum Degenerate Gas group at LENS for discussions.

\end{document}